\documentclass[11pt]{article}

\usepackage{amsmath,amsfonts,amssymb}
\usepackage{graphicx}
\usepackage{secdot}

\author{\footnotesize Warner A. Miller$^1$, Shahabeddin Mostafanazhad Aslmarand$^1$, Paul M. Alsing$^2$\and \footnotesize Verinder S. Rana${}^3$}
\date{\footnotesize %
   \itshape{ $^1$Department of Physics, Florida Atlantic University, Boca Raton, FL, 33431, USA}\\%
    $^2$Air Force Research Laboratory, Information Dire
   ctorate, Rome, NY, 13441, USA\\%
    $^3$Naval Information Warfare Center Pacific (NIWC PAC), San Diego, CA 92152, USA
}
\begin{document}

\title{Geometric Measures of Information
for Quantum State Characterization}

\maketitle

\begin{abstract}
We analyze the geometry of a joint distribution over a set of discrete random variables.  We briefly review Shannon's entropy, conditional entropy, mutual information and conditional mutual information. We review the entropic information distance formula of Rokhlin and Rajski.  We then define an analogous information area.  We motivate this definition and discuss its properties. We extend this definition to higher-dimensional volumes. We briefly discuss the potential utility for these geometric measures in quantum information processing.
\end{abstract}


\section{Entropy and Mutual Information}
\label{sec:intro}

\begin{figure}[h]
 \centering
 \includegraphics[width=4in]{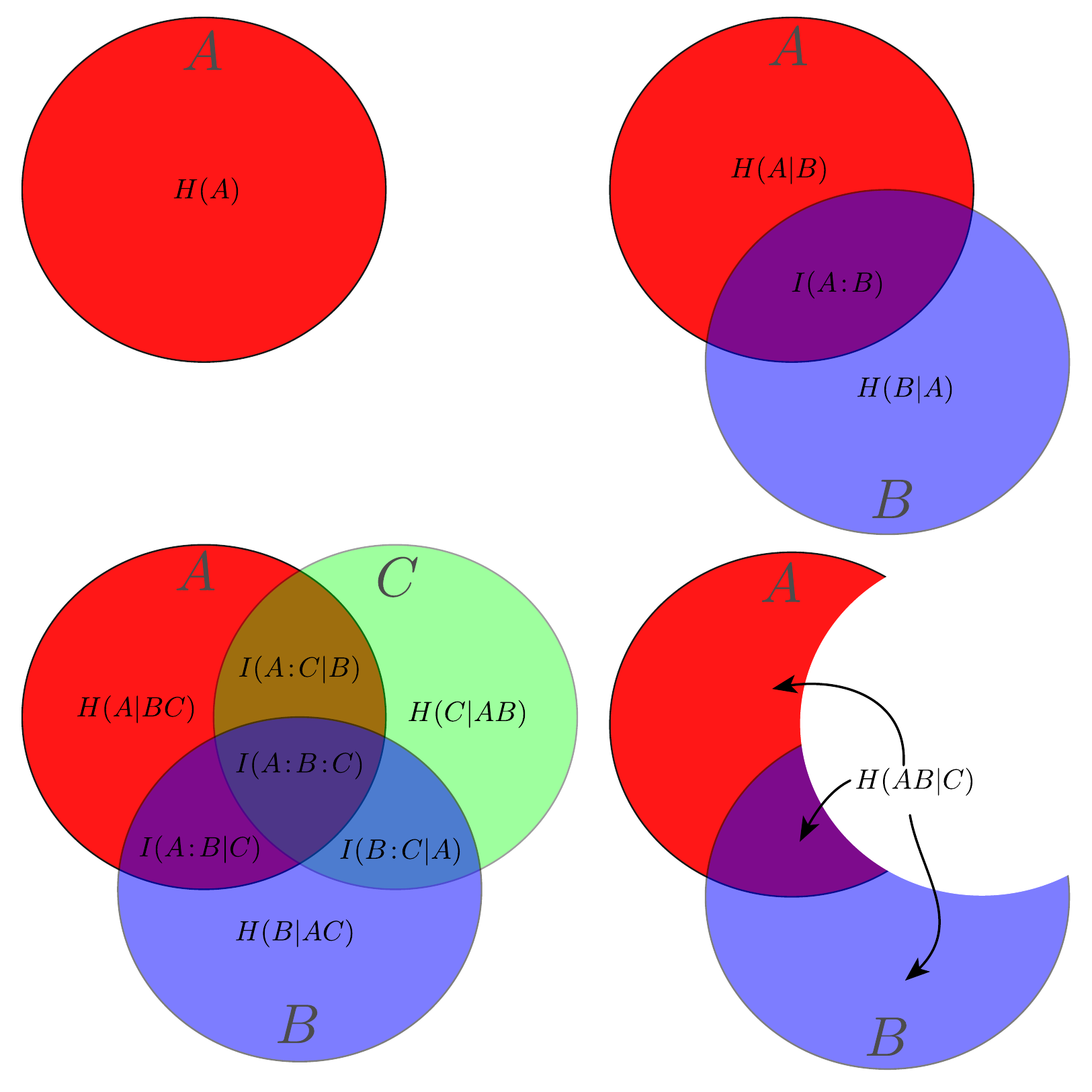}
 \label{fig:venn}
 \caption{Venn diagram of entropies for three random variables, $A$, $B$ and $C$.}
 \end{figure}

Probability measures the uncertainty in a single measured or observed event, while entropy measures the uncertainty in a collection of events.  In this section, we briefly review the Shannon--based entropy, conditional entropy, mutual information and conditional mutual information for three discrete random variables, $A$, $B$ and $C$ as illustrated in the Venn diagram in Fig.~\ref{fig:venn}.  The Venn diagram provides a convenient geometric representation of the entropy and mutual information measures. We introduce the entropy primarily to define our notation; however, we use the entropies to define the information distance, area, volume and higher-dimensional generalizations in the following  Secs.~\ref{sec:d}--\ref{sec:v}. In Sec.~\ref{sec:fini} we summarize recent work showing the potential utility of these measures have in quantum mechanics by following the It-from-Bit  formalism first introduced by John Archibald Wheeler on the quantum.\cite{Wheeler:1990} In Sec.~\ref{sec:fini} we briefly discuss how these the entropic $n$--volumes can be used to define a measure of quantum--correlation  that is scalable to higher-dimensional multipartite states.\cite{Aslmarand:2019,Aslmarand:2019b,Miller:2018} 

It is well accepted that quantum entanglement is the key resource for quantum computing,\cite{Preskill:2012} and this has led to many studies on this topic.\cite{Peres:1996,Wootters:1998,Terhal:2000,Rungta:2003,HHHH:2009,Huber:2010,Rulli:2011,Eltschka:2012}   We recently proposed a novel information--geometric measure of quantum correlation, that we refer to as quantum reactivity $\mathcal R$.\cite{Aslmarand:2019,Aslmarand:2019b}  It is similar to the reactivity in chemistry in that it is a ratio of surface area to volume.  However, our areas and volumes are entropic in nature and are defined on the outcomes of measurements  (a ``space of measurements'') from an ensemble of identically--prepared quantum systems.  We found that $\mathcal R$ is monotonic, and when averaged over the space of measurements,  this reactivity $\overline{\mathcal R}$ is invariant under local unitary operations on subsystems of the quantum state.\cite{Aslmarand:2019b}  Although it may be rather surprising that one could even consider a  Shannon-entropy approach based on measurements, nevertheless Schumacher showed that this was indeed possible,\cite{Schumacher:1991} and we have reproduced this in the laboratory.\cite{Razaei:2019} We feel it is important to find such a scalable and well--defined measure. Once found, one can consider approximations, computational efficiency and experimental feasibility.  Measures of multipartite entanglement is a difficult and unresolved problem, and perhaps its solution may involve a novel geometric approach that will provide new clues for such a metric?  Geometry is often a  guide for such solutions, and in this spirit, we define and examine information area, volume and $n$--volumes in this manuscript. 

\subsection{Entropy of a Single Random Variable, $A$}
\label{sec:A}

Consider a discrete random variable, $A$, that takes on $s$ possible values, $a_i$, each with probability $p_{a_i}$, where 
\begin{equation}
p\left( A\!=\!a_i \right) = p_{a_i}, \ \  \forall i\in\{1,2, \ldots s\}. 
\end{equation}
Probability measures the uncertainty of a single event, entropy measures the uncertainty in a collection of events, i.e. a random variable $A$. The Shannon entropy of $A$ is
\begin{equation}
H\left(A\right) = -\sum_{i=1}^{s} p_{a_i} \log p_{a_i},
\end{equation}
where 
\begin{equation}
\sum_{i=1}^{s} p_{a_i} = 1.
\end{equation}
In this manuscript we use the logarithm base two, and we use two notations for entropy, $H_A\equiv H(A)$. 

The entropy measures the uncertainty in $A$ and is bounded
\begin{equation} 
0 \le H\left( A\right) \le \log s.
\end{equation}
It takes on its minimum value when there is no uncertainty in $A$, i.e. when $p_{a_j}\!=\!1$ for some $j$. It assues its maximal bound for the most uncertain state or uniform distribution, $p_{a_i}\!=1/s$ for all values $j$.  The entropy $H\left(A\right)$ is depicted as the solid red disc in the upper left of the Venn diagrams illustrated in Fig.~\ref{fig:venn}.

\subsection{Conditional Entropy and Mutual Information of Two Random Variables, $A$ and $B$}

We now outline the entropies of a pair of discrete random variables $A$ and $B$ using the same notation as we used in Sec.~\ref{sec:A}.  These are illustrated in the upper right hand of Fig.~\ref{fig:venn}. Ordinarily, these two random variables will be described completely by their normalized joint probability distribution $\rho_{AB}$ with probabilities
\begin{equation}
P\left(A\!=\!a_i,B\!=\!b_j\right) = p_{a_ib_j}, \ \ \forall i,j\in\left\{1,2,\ldots,s\right\},
\end{equation}
and 
\begin{equation}
\sum_{i,j=1}^{s} p_{a_ib_j} = 1.
\end{equation}
Their joint entropy is given by a double summation, 
\begin{equation}
H\left(AB\right) = -\sum_{i,j=1}^{s} p_{a_ib_j} \log p_{a_ib_j}.
\end{equation}
This joint entropy is bounded from above and below, 
\begin{equation}
\label{eq:bounds2} 
0 \le H\left( AB\right) \le H\left(A\right) + H\left(B\right) = 2 \log s,
\end{equation}
where it takes on its minimum value when there is no uncertainty in the joint distribution, i.e. when $p_{a_ib_j}\!=\!1$ for some $i$ and $j$ and all other probabilities are zero. In which case $H(AB)=H(A)=H(B)=0$.  It assumes its maximal bound when the variables $A$ and $B$ are mutually independent (disjoint $A\cap B=0$) with each being a uniform distribution, $p_{a_i}= p_{b_j}\!=1/s$ for all values $i$ and $j$.  In which case, $H(AB)=H(A)+H(B)$ and $H(A)=H(B)=\log s$. The joint entropy $H\left(AB\right)$ is depicted as the union of the solid red disc and the semi-opaque blue disc in the upper right of the Venn diagrams illustrated in Fig.~\ref{fig:venn}. It is the sum of the three areas; (1) the red crescent or conditional entropy $H\left(A|B\right)$, (2) the blue crescent conditional entropy $H\left(B|A\right)$ and (3) the purple overlapping area given by the mutual information $I\left(A\!:\!B\right)$,
\begin{equation}
H\left(AB\right) = \underbrace{H\left(A|B\right)}_{red\ crescent}+\underbrace{H\left(B|A\right)}_{blue\ crescent}+\underbrace{I\left(A\!:\!B\right)}_{purple\ overlab}.
\end{equation}
The entropic Venn diagram is quite useful for such formulas.
Each of these expressions can also be expressed as a function of probabilities, the joint entropy is 
\begin{equation}
\label{eq:ceagb}
H\left(A|B\right) =\sum_{i,j} p(A=a_i,B=b_j) \log{p(A=a_i|B=b_j)}, 
\end{equation}
and the mutual information is 
\begin{equation}
\label{eq:miab}
I\left(A\!:\!B\right) =\sum_{i,j} p(A=a_i,B=b_i) 
\log{
\frac{p(A=a_i,B=b_i)}{p(A=a_i)p(B=b_j)}
}.
\end{equation}
Here the discrete random variable $B$ is obtained by tracing out over all values of $A$,
\begin{equation}
p(B=b_j) = \sum_i p(A=a_i, B=b_j),
\end{equation}
and the conditional probability is the ratio
\begin{equation}
p(A=a_i|B=b_i) =\frac{p(A=a_i,B=b_i)}{p(B=b_j)}.
\end{equation}

\subsection{Conditional Entropy, Mutual Information and  Conditional Mutual Information of Three Random Variables, $A$, $B$ and $C$}

The entropies and mutual information we outlined in the last subsection can be extended easily to higher dimensions.  We apply this to three random variables in this subsection.  The joint entropy of three discrete random variables $A$, $B$ and $C$,  is given by the triple summation, 
\begin{equation}
H\left(ABC\right) = -\sum_{i,j,k=1}^{s} p_{a_ib_jc_k} \log p_{a_ib_jc_k}.
\end{equation}
Just as the bounds we calculated for two random variables in Eq.~\ref{eq:bounds2}, the joint entropy for three variables is also bounded from above and below, 
\begin{equation}
\label{eq:bounds3} 
0 \le H\left( ABC\right) \le H\left(A\right) + H\left(B\right) + H\left(C\right)= 3 \log s.
\end{equation}
We saw the utility of Venn diagram in Fig.~\ref{fig:venn} to derive entropic relations, the chain rule for entropy is also useful for relating conditional entropies with joint entropies,
\begin{equation}
\label{eq:chain3} 
H(ABC) = H(A) + H(B|A) +H(C|AB),
\end{equation}
 from which we find the conditional entropy
\begin{eqnarray}
\label{eq:ceabc}
H\left(C|AB\right) &=& H\left(ABC \right) - H\left(A\right)-H\left(B|A\right)\\
&=& H\left(ABC \right) - H\left(B\right)-H\left(A|B\right)\\
&=& H\left(ABC \right) - H\left(AB\right).
\end{eqnarray}
Here we also used the chain rule for two variables, $H(AB)=H(A)+H(B|A)$. The chain rule is straightforwardly generalizable to higher dimensions.
 
We emphasize again that this relation (Eq.~\ref{eq:chain3}) as well as the other entropic relations in this manuscript can be easily verified and obtained by inspection from the Venn diagrams in Fig.~\ref{fig:venn}.  For example, the conditional entropy is easily seen to be
\begin{eqnarray}
H\left(AB|C\right) &=&  H\left(B|AC\right) + H\left( A|C\right)\\
&=& H(ABC) -H(C). 
\end{eqnarray}
Such expressions in three and higher dimensions can also be expressed in terms of probabilities similar to Eqns.~\ref{eq:ceagb}--\ref{eq:miab}. In particular,  the mutual information for three variables is similar to the mutual information for two variables in Eq.~\ref{eq:miab}, where
\begin{equation}
\label{miabc}
I\left(A\!:\!B\!:\!C\right) =\sum_{i,j,k=1}^s p(A=a_i,B=b_i,C=c_k) 
\log{
\frac{p(A=a_i,B=b_j,C=c_k)}{p(A=a_i)p(B=b_j)p(C=C_k)}
}.
\end{equation}
The joint mutual information $I\left(A\!:\!B\!:\!C\right)$ of the three variables is represented by the bowed--out triangular area of intersection ($A\cap B\cap C$) of the three discs as illustrated in the lower left  of Fig.~\ref{fig:venn}.
\begin{eqnarray}
I\left(A\!:\!B\!:\!C\right) &=& I\left(A\!:\!B\right)  - I\left(A\!:\!B | C\right) \\
&=&  H\left(A B C\right) - H\left(B|A\right) - H\left(A|C\right) - H\left(C|B\right) \\
&=& H\left(A B C\right) - H\left(B C\right) - H\left(A C\right) - H\left(A B\right) + \nonumber \\
&& H\left(A\right)  + H\left(B\right) + H\left(C\right).
\end{eqnarray}
From the Venn diagram we see that the conditional mutual information can be expressed in terms conditional and joint entropies,  
\begin{eqnarray}
I\left( A\!:\!B|C\right) &=& H\left(ABC\right) -H\left(B|AC\right) -H\left(A|BC\right) - H\left(C\right)\\
&=& - H(ABC) +H(AC)+H(BC)  - H(C).
\end{eqnarray}

\section{The Rokhlin and Rajski Information Distance}
\label{sec:d}

Rokhlin and Rajsk defined an information metric based on the Shannon entropy, where 
\begin{eqnarray}
 \label{eq:D}
 D_{AB} &=& H\left(A|B\right) + H\left(B|A\right) \nonumber \\ 
 &=& 2H\left(AB\right)-H\left(A\right)-H\left(B\right),
 \end{eqnarray}
where $H_{A|B} $ is conditional entropy of $A$ given $B$ defined in Eq.~\ref{eq:ceagb}.\cite{Rokhlin:1967,Rajski:1961,Zurek:1989,Bennett:1998}

Using the properties of Shannon entropy it's possible to show that this measure of distance satisfies all three properties of a metric,
\begin{itemize}
\item[a)] It is constructed to be symmetric, $D_{AB} = D_{BA}$,
\item[b)] It obeys the triangle inequality, $D_{AB}\geq D_{AC}+ D_{CB}$, and 
\item[c)] It is non-negative, $D_{AB}\geq 0$, and equal to zero when $A$“=”$B$.
\end{itemize}

\section{An Information Area and Its Properties}
\label{sec:a}

In addition to the entropic information distance, we can analogously assign an entropic  information area.  An information area, ${\mathcal A}_{ABC}$,  defined over three discrete random variables $A$, $B$ and $C$ should obey the following minimum properties:
\begin{itemize}
\item[a)] It should be two-dimensional and have dimensions $bit^2$. Since the unit of entropy is bits (i.e. $\left[H\right] = bit$) then the area should be quadratic in a measure of entropy;
\item[b)] It should be symmetric under the interchange of any two random variables;
\item[c)] It should vanish when any two random variables are maximally correlated, i.e. when $A$`='$B$; 
\item[d)] It should be a function of the tripartite correlation, i.e. it should functionally depend, in some way to the joint entropy $H\left(ABC\right)$;
\item[e)] It should be bounded above when all three random variables are uncorrelated, and bounded below when the three random variables are maximally correlated ($A$`='$B$`='$C$), in which case ${\mathcal A}_{ABC} \ge 0$ (positivity);
\end{itemize}
An area satisfying these five properties and a generalization of Eq.~\ref{eq:D} was developed in discussions at the 1989 Santa Fe Institute conference on ``Complexity, Entropy, and the Physics of Information'' between Caves, Kheyfets, Lloyd, Miller, Schumacher and Wootters \cite{QIG:1990} in order to explore possible information-based descriptions of spacetime in general relativity.  More recently we applied these entropic areas and volumes to quantum states for a novel entanglement measure.\cite{Miller:2018}.  The area for a joint probability distribution over three discrete random variables was defined by generalizing Eq.~\ref{eq:D} as  
\begin{equation}
\label{eq:A}
{\mathcal A}_{ABC} = H_{A|BC}H_{B|CA}+H_{B|CA}H_{C|AB}+H_{C|AB}H_{A|BC},
\end{equation}
where we introduced the notation for entropy, e.g. $H_{A|BC}\equiv H(A|BC)$. This notation will be used for  the remainder of the manuscript.
This can also be written in terms of joint entropies using Eq.~\ref{eq:ceabc}, 
\begin{eqnarray}
\label{eq:Aje}
{\mathcal A}_{ABC} &=& 3 H_{ABC}^2 -2 \left[H_{AB}+H_{AC}+H_{BC}\right] H_{ABC} + \nonumber \\
&&  \left(H_{AB}H_{BC}+H_{AB}H_{AC}+H_{AC}H_{BC}\right).
\end{eqnarray}

The definition of the area based on the five conditions is not unique.  In particular, the area 
\begin{equation}
\label{eq:contrived}
\frac{ {\mathcal A}_{ABC}+\bigtriangleup_{ABC}}{2}
\end{equation}
is an acceptable area.  Here
\begin{eqnarray}
\bigtriangleup^2_{ABC} &=& 
\frac{1}{4} \left[ 
\left( H_A+H_B+H_C\right)
\left( H_A-H_B+H_C\right)  \right. \nonumber \\
&&\left.  \left( H_A+H_B-H_C \right)
\left(- H_A+H_B+H_C \right)
\right]
\end{eqnarray} 
is the usual Euclidean triangle area as a function of the lengths of its three edges using the entropic information distance in Eq.~\ref{eq:D}.  It is perhaps worth noting that he Euclidean area, $\bigtriangleup^2_{ABC}$ ,  alone is insensitive to correlations shared by all three random variables; however, the combination of the two areas is a function of the tripartite correlations.  Eq.~\ref{eq:contrived} is rather contrived, but is useful to prove non uniqueness of area based for a non-trivial example.  We forward the definition of entropic area in Eq.~\ref{eq:A} as the most reasonable definition that we can imagine.  It is a hardly unambiguous but reasonable generalization of the information distance formulae in Eq.~\ref{eq:D}. 

\section{An Information Volume:  Its Properties and Generalization to Higher Dimensions}
\label{sec:v}

An analogous information volume for a tetrahedron generalizing Eq.~\ref{eq:A} is \cite{QIG:1990,Aslmarand:2019,Miller:2018}
\begin{equation}
\label{eq:V}
\begin{array}{ll}
{\mathcal V}_{ABCD}& := H_{A|BCD}H_{B|CDA}H_{C|DAB} +H_{B|CDA}H_{C|DAB}H_{D|ABC}\\
&\ \ +H_{C|DAB}H_{D|ABC}H_{A|CDB}+H_{D|ABC}H_{A|BCD}H_{B|CDA}.
\end{array}
\end{equation}
This too satisfies the analogous five criteria:
\begin{itemize}
\item[a)] It should be three-dimensional and have dimensions $bit^3$. Since the unit of entropy is bits (i.e. $\left[H\right] = bit$) then the volume should be cubic in a measure of entropy;
\item[b)] It should be symmetric under the interchange of any two random variables;
\item[c)] It should vanish when any two random variables are maximally correlated, i.e. when $A$`='$B$;
\item[d)] It should be a function of the quadripartite correlation, i.e. it should functionally depend, in some way to the joint entropy $H\left(ABCD\right)$;
\item[e)] It should be bounded above when all three random variables are uncorrelated, and bounded below when the three random variables are maximally correlated ($A$`='$B$`='$C$`='$D$), in which case ${\mathcal A}_{ABCD} \ge 0$ (positivity);
\end{itemize}

Consider  a discrete multipartite state, $\mathcal S$ consisting of $d$ random variables
\begin{equation}
{}^{^{(d)}}\!{\mathcal S}=\left\{A_1,A_2, \ldots, A_d\right\}
\end{equation}
that form a $(d-1)$--dimensional complete graph.  
The  volume of this  $(d-1)$--dimensional simplex that that shares the structure of Eq.~\ref{eq:V} is  
\begin{eqnarray}
\label{eq:nV}
{}^{^{(d\!-\!1)}}\!{\mathcal V}_{A_1A_2\ldots A_d} &:=& 
\displaystyle
\sum_{a_1,a_2,\ldots a_d=A_1}^{A_d} 
\left(  \frac{1+\epsilon_{a_1a_2\ldots a_d}}{2}  \right) \nonumber \\
&&\underbrace{
H_{A_1|a_2\ldots A_{d}}H_{A_2|a_1A_3\ldots A_{d}} \ldots H_{A_{d-1}|A_1A_2\ldots A_{d\!-\!2} A_{d}}
}_{
\hbox{\tiny product of}\ d\!-\!1\ \hbox{\tiny conditional entropies}}, 
\end{eqnarray}
where $\epsilon_{a_1a_2\ldots a_d}$ is the Levi-Civita tensor with $d$ indices.

\section{Conclusion}
\label{sec:fini}
We defined and motivated the definitions for entropic area (Eqn.~\ref{eq:A}), volume (Eqn.~\ref{eq:V}) and their higher--dimensional generalizations in (Eqn.~\ref{eq:nV}).  Our definitions of these entropic geometry formulae were motivated by Wheeler's It--from--Bit  research program.\cite{Wheeler:1990}   We developed these entropic geometric measures to explore the geometric structure of quantum states, and to find  a measure of quantum entanglement.\cite{Aslmarand:2019,Miller:2018}  Entanglement measures based on relative entropy have been introduced earlier and perhaps it is worth revisiting these using the geometric insight we provided.\cite{Wei:2008,Bravyi:2003}  Geometry is often a valuable guide for scientific discovery. There may be many more applications; however, we discuss here our application of these entropic geometry measures to quantum mechanics. 

One of the central problems in quantum information processing today is a proper and scalable entanglement measure for high-dimensional multipartite quantum states.  By proper, we mean a measure that is  (1) invariant under local unitary transformations on a subsystem of the state, and (2) non--increasing under LOCC, i.~e.  monotonically increasing in the amount of quantum correlation. Quantum correlation and entanglement is recognized as the key resource for quantum information processing.\cite{Preskill:2012}  This is a difficult and unresolved problem for multipartite quantum systems. 

As discussed in Sec.~\ref{sec:intro}, one may have pause to question the applicability of geometric measures based on Shannon's entropy to quantum states and their quantum correlations as it ordinarily would not be invariant under local unitary transformations,  much less be a monotonically-increasing function of the amount of quantum correlation.  Nevertheless, we have recently shown the volume formulae introduced in this paper appear to yield a reasonable measure for quantum correlation satisfying accepted  conditions.\cite{Aslmarand:2019,Aslmarand:2019b}  In particular, we have found recently that quantum reactivity appears to be a viable measure of quantum correlations for a multipartite state.\cite{Aslmarand:2019}   The term reactivity, as commonly used for chemical reactions, is a ratio of surface area to volume.  This can be defined for an $d$-partite quantum system using the entropic volume formulae introduced in this manuscript as a weighted sum over the ``space of measurements,'' 
\begin{equation}
\label{eq:reactivity}
{\mathcal R} := \frac{
\langle {}^{^{(d\!-\!2)}}\!\!{\mathcal A} \rangle_{\mathcal M}
}{
\langle {}^{^{(d\!-\!1)}}\!{\mathcal V} \rangle_{\mathcal M}
}.
\end{equation}
Here we defined ${}^{^{(d\!-\!2)}}\!\!{\mathcal A}:=\!\!{}^{^{(d\!-\!2)}}\!{\mathcal V}$, and we dropped  the subscripts for clarity. The integral in the weighted sum is over all measurements.\cite{Aslmarand:2019}

This geometric approach based on Shannon's entropy of the quantum measurement outcomes appears to provide a viable measure of quantum correlation, and is a generalization of Schumacher's approach.\cite{Schumacher:1991}  It is scalable in that it can be generalized to an arbitrary large multipartite state in general position. However, the computational resources needed to compute the reactivity is formidable. One has at a minimum to calculate the joint entropy of all the $d$ qubits of a multipartite state, and this requires computing an exponential number of terms in $d$.  Nevertheless we ask,  ``Is  possible, for a broad class of multipartite states, to achieve exponential convergence in fidelity to the state after a finite number ($n>>d$) of measurements?''\cite{Vedral:1997}  Perhaps the integral in Eq.~\ref{eq:reactivity} can be approximately replaced with a sum over a finite number of measurements (i.e. detector settings)?  Since quantum reactivity satisfies the major conditions required for a measure of quantum correlation, can this geometric technique be used to reformulate some unsolved problems in measures of entanglement?

Although we discussed our current interest in entropic volumes for measures of quantum correlation, we can very well imagine that these volume formulas might find utility in other fields of mathematics, science or engineering. Ukichiro Nakaya (physicist) refers to a snowflake as a letter from the sky, and classified over forty classes of snow flakes that were used to correlate to atmospheric conditions.\cite{David:2003}  Similarly, in a poem by Frank, "A diamond is a letter from the depth of the earth."\cite{Yonezawa:1983}  Perhaps one can say:  {\em Quantum entanglement is a letter from the universe.}

\section{Acknowledgments}
PMA and WAM would like thank support from the Air Force Office of Scientific Research (AFOSR).  WAM  research was supported under AFOSR/AOARD grant \#FA2386-17-1-4070. WAM wished to thank the Griffiss Institute and AFRL/RI for support under the Visiting Faculty Research Program.  Any opinions, findings, conclusions or recommendations expressed in this material are those of the author(s) and do not necessarily reflect the views of AFRL.

%
 \section*{Conflict of interest}
The authors declare that they have no conflict of interest.

\bibliographystyle{plain}

\bibliography{qig5-2019}  

%
%

\end{document}